\documentclass{article}
\usepackage{waspaa21,amsmath,graphicx,url,times}
\usepackage[table,xcdraw]{xcolor}
\usepackage{boldline}
\usepackage{enumitem}
\usepackage{setspace}
\usepackage{slantsc}
\usepackage{multirow}
\usepackage{dsfont}
\usepackage{color}
\usepackage{arydshln}
\usepackage{booktabs}
\usepackage{hyperref}
\usepackage{enumitem}
\usepackage{boldline}
\usepackage{enumitem}
\usepackage{setspace}
\usepackage{slantsc}
\usepackage[export]{adjustbox}
\usepackage[lofdepth,lotdepth]{subfig}
\usepackage[tracking=true]{microtype}
\usepackage{cite}
\usepackage{xspace}

\title{DPLM: A Deep Perceptual Spatial-Audio Localization Metric}

\name{\narrowstyle Pranay Manocha,$^{1 \star}$ \thanks{$^{\star}$ \footnotesize{This work was performed during an internship at Facebook Reality Labs Research}}
      Anurag Kumar,$^{2}$
      Buye Xu,$^{2}$
      Anjali Menon,$^{2}$
      Israel D. Gebru,$^{2}$
      Vamsi K. Ithapu,$^{2}$
      Paul Calamia$^{2}$ \vspace{-0.08in}}

\address{$^1$ Dept. of Computer Science, Princeton University, Princeton, NJ USA\\ $^2$ Facebook Reality Labs Research, Redmond, WA USA\\
          pmanocha@cs.princeton.edu, \{anuragkr, xub, aimenon, idgebru, ithapu, pcalamia\}@fb.com
           \\
\vspace{-0.3in}
}

\DeclareUnicodeCharacter{2212}{-}

\newcommand{\ignorethis } [1] {}

\newcommand{\eg         }     {{e.g.}}

\newcommand{\Reals      }     {{\textrm{I\kern-0.18em R}}}

\newcommand{\change     } [1] {\mbox{{\footnotesize $\Delta$} \kern-3pt}#1}

\newlength{\w}

\newcommand\narrowstyle{\SetTracking{encoding=*}{-50}\lsstyle}

\newcommand{\PESQ}   {PESQ}

\newcommand{\POLQA}  {POLQA}

\newcommand{\DPAM}   {DPAM}
\newcommand{\CDPAM}   {CDPAM}

\newcommand{\OURS}   {DPLM}
\newcommand{\BAMQ}   {BAMQ}
\newcommand{\TASNET}   {TasNet}
\newcommand{\SAGRNN}   {SAGRNN}

\newcommand{\Fmodel}   {{{F}}}

\begin{document}

\ninept
\maketitle

\begin{sloppy} 
\vspace{-1in}
\begin{abstract} 
\vspace{-0.02in}
Subjective evaluations are critical for assessing the perceptual realism of sounds in audio-synthesis driven technologies like 
augmented and virtual reality.~However, they are challenging to set up, fatiguing for users, and expensive.~In this work, we tackle the problem of capturing the perceptual characteristics of localizing sounds.~Specifically, we propose a framework for building a general-purpose quality metric to assess spatial localization differences between two binaural recordings.~We model localization similarity by utilizing activation-level distances from deep networks trained for direction of arrival (DOA) estimation.~Our proposed metric (\OURS) outperforms baseline metrics on correlation with subjective ratings on a diverse set of datasets, even without the benefit of any human-labeled training data.


\end{abstract}

\vspace{-0.08in}

\vspace{-0.02in}
\section{Introduction}
\label{sec:intro}
\vspace{-0.02in}
Perceptually realistic audio and sound-processing systems are vital for immersive multi-sensory 
technologies like Augmented and Virtual Reality (AR and VR) experiences.~Such processing systems may include the synthesis of realistic-sounding audio, accurate spatial presentation of 3D virtual sounds, or, more broadly, high-quality rendering of virtual audio.~Sound-quality evaluation tests are critical because they validate the resulting user experience and also provide necessary user feedback that drives the synthesis pipeline.~Nonetheless, the inherent subjectivity in designing such tests makes it difficult to develop multi-purpose evaluation mechanisms that take various aspects of sound quality into account.


In this paper, we focus on the problem of accurate {\it binaural} presentation of sound sources in the far-field.~Such presentation drives the perceptual quality of spatial audio in AR and VR.
~The ideal approach to characterizing binaural sound-source localization is to first synthesize 
the necessary sound signals and then perform a listening test via user studies.~This process may be repeated hundreds of times for different combinations of source locations, which is costly and time-consuming.
Further, the majority of recent audio-processing algorithms are machine learning (or deep learning) driven and rely on large labeled datasets.
This makes such exhaustive listening tests impractical.~The widespread use of such end-to-end systems driven by neural networks also necessitates the 
design of testing models that are {\it differentiable}, \emph{i.e.}, one can back-propagate errors from listening tests directly to the inputs.~As a result, an efficient and robust objective metric that can effectively substitute for a subjective listening test is required.

Several research works have proposed objective metrics based on binaural cues like Interaural Level Differences (ILD), Time Differences (ITD) and Cross-Correlation (IACC)\cite{kampf2010standardization,flessner2017assessment,seo2013perceptual,takanen2012binaural} to evaluate spatial audio quality.~However, they suffer from various general drawbacks.~First, they are sensitive to background noise - hindering their usage in diverse realistic sound synthesis scenarios.~Second, they typically work well only under anechoic conditions and are not accurate in reverberant environments.~Third, they don't take into account complex scenes with multiple sources.~Lastly, they assume that the two binaural signals to be compared are time-aligned and of equal length, which is not always the case.~Researchers have proposed identifying the number of participating sources before using binaural cues~\cite{schafer2013extension,seo2013perceptual}, which addresses the multiple-source aspect, but the rest  of  the drawbacks remain. 

On the other hand, one may consider adapting existing objective assessment metrics for quality of monaural signals such as 
\PESQ\,~\cite{rix2001perceptual}, \POLQA\,~\cite{beerends2013perceptual}, \DPAM\,~\cite{manocha2020differentiable} and \CDPAM\,~\cite{manocha2021cdpam} for this task. 
However, since these metrics only focus on perceived quality rather than spatialization, their utility for multi-channel signals remains limited~\cite{kampf2010standardization,conetta2015spatial}. 
Some researchers have recently looked at problem-specific (\eg\, audio-coding) models for objective assessment of binaural audio quality~\cite{seo2013perceptual,takanen2012binaural,schafer2013extension,delgado2019objective,narbutt2020ambiqual}. 
Delgado et al.~\cite{delgado2019objective} address the specific use-case of collapsing the stereo image to the center at low bitrates, 
whereas Narbutt et al.~\cite{narbutt2020ambiqual} compare ambisonic signals for audio codecs. 
These models are \emph{non-differentiable}, though, and cannot be directly leveraged as a training objective for deep networks. 
Also, they require human-annotated datasets for training or calibration, which often are not publicly available. 

\begin{figure*}[h!]
\vspace{-0.23in}
\centering
\setlength{\w}{0.45\textwidth}
\setlength{\tabcolsep}{8pt}
\begin{tabular}{cc}
\includegraphics[width=0.50\textwidth]{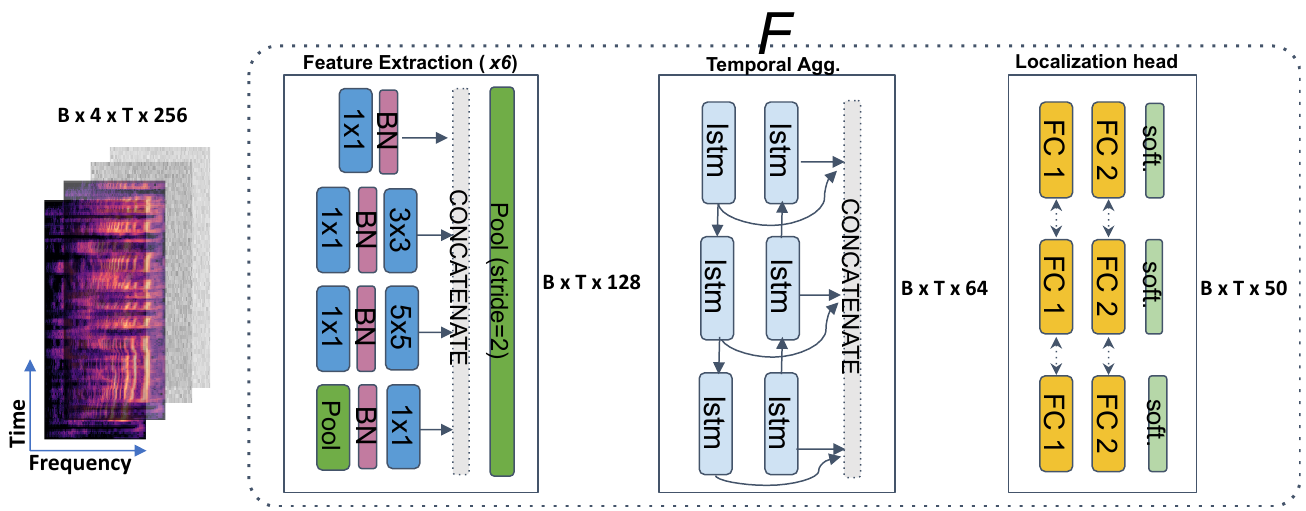} &
\includegraphics[width=0.425\textwidth]{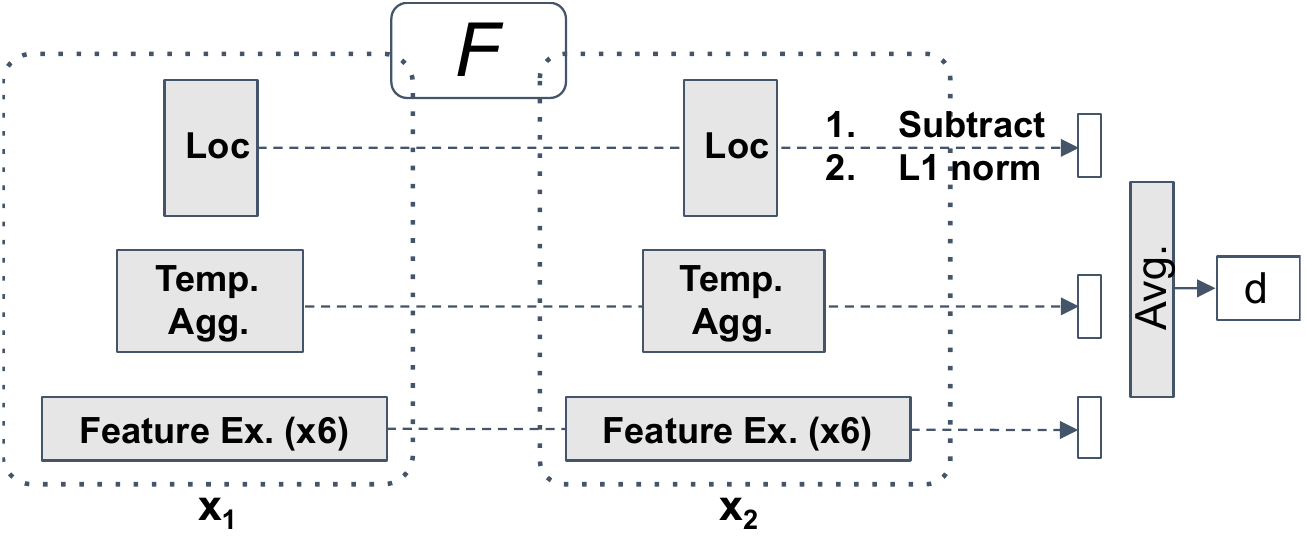}\\

\begin{minipage}{\w}
\centering
\vspace{-0.1in}
{\footnotesize \bf(a)}

\end{minipage} &
\begin{minipage}{\w}
\centering
\vspace{-0.1in}
{\footnotesize \bf(b)} 
\end{minipage}
\end{tabular}
\vspace{-3.5ex}
\caption{\textbf{\OURS\, architecture}: (a)~We first train a source localization model $\Fmodel$ to estimate framewise \emph{DOA}, then (b)~use the extracted \emph{deep-features} to compute a distance $D(x_1,x_2)=d$ between the two binaural signals $x_1$ and $x_2$}
\label{model}
\vspace{-3ex}
\label{model_architecture}
\end{figure*}

We propose a framework for learning a binaural-audio similarity metric that addresses some of these issues. 
Specifically, we propose \OURS\,: a full-reference \underline{d}eep \underline{p}erceptual spatial audio \underline{l}ocalization \underline{m}etric  that evaluates the similarity of binaural presentations in terms of localization.~We begin by building binaural \emph{direction-of-arrival (DOA)} deep network models that act as surrogates for localization.~Given two different inputs, a simple difference of the model output layer representations between these inputs can, 
in principle, represent a localization metric.~However, DOA estimations are typically sensitive to noise, reverberation, and sound source characteristics, which strongly affect the accuracy of localization assessment.~Instead, we compute \emph{deep-feature distances}~\cite{zhang2018unreasonable} between the full-feature activation stacks of the DOA model to assess localization similarity between sound sources.~To further improve robustness, we train DOA models with carefully designed input perturbations as data augmentations that mimic realistic environments.~We show that, even in the absence of explicit perceptual training, these distances correlate well with human perceptual judgments (both via objective and subjective tests).~We also show that the resulting metric generalizes even for distinct (yet related) tasks such as audio codecs, binaural reproduction from mono or multi-channel signals, etc.~And since the metric is based on a deep network, it is \emph{differentiable}, and can be directly leveraged as a training objective for localization and related audio and sound-source synthesis tasks.




\section{The DPLM Metric} \label{sec:dplm}

In this section we describe the \OURS\, framework (Fig~\ref{model}).~Given two binaural signals denoted by $x_\mathtt{1}$ and $x_\mathtt{2}$, our goal is to compute a distance function $D(x_\mathtt{1},x_\mathtt{2})$ that precisely characterizes the localization similarity between them.~The distance function is designed to be \emph{non-negative} and \emph{monotonic}, thereby making it a pseudo-metric (we do not impose triangle or associative properties).

\subsection{DOA Model Strategies} \label{doa_models}

We begin by building a source-localization model that predicts the DOA of a given sound source.
~Given a binaural input, the DOA model processes both the magnitude and phase spectrograms of the two channels, and systematically outputs a framewise location estimate of the source.~We evaluate two variants to ensure generalization of the proposed framework for realistic sound sources: \noindent a {\bf \emph{Static-Source}} model for recordings with a fixed source in the scene, and \noindent a {\bf \emph{Moving-Source}} model where the source can move smoothly in the scene.~Note that the moving-source model would result in a finer resolution estimate of DOA.

\subsection{Architecture} \label{training_architecture}

\OURS\ comprises three components (Fig~\ref{model_architecture}a): 
a \emph{feature-extraction} block, a \emph{temporal aggregation} block and a task-specific \emph{localization block}.~Both DOA variants have the same model architecture for fair comparison.~
The feature-extraction block maps the input features to an embedding maintaining the temporal structure of the signal. 
This embedding is then processed by temporal aggregation models to learn long-term dependencies.
~The resulting hidden representations are then fed to a task-specific localization head 
that outputs a location embedding per frame.

For the feature-extraction block, we evaluated a variety of different convolutional network building blocks, including a basic conv-batchnorm-maxpool block, ResNet, Squeeze-and-Excitation and Inception~\cite{szegedy2015going}. 
Inception resulted in best features, and so, we do not discuss the other structures in this paper. 
The 6-block Inception network consists of 64 conv filters, followed by 1x1, 3x3 and 5x5 filters, 
leading to 3x3 max-pooling, and a 1x2 maxpool along the frequency dimension.~For the temporal aggregation block, we evaluated both Long Short-Term Memory networks (LSTMs) and Temporal Convolutional Networks (TCNs). 
LSTMs generalized better to unseen rooms and subjects. 
We used 2 bi-directional LSTM layers that output an embedding of size 64 for each time-frame.

We pose the localization task as a simple classification problem.~We divide the azimuth plane ($\in(-180^{\circ}, 180^{\circ})$) into 50 equally spaced bins, and the elevation plane ($\in(-90^{\circ}, 90^{\circ})$) into 25 equally spaced bins. 
However, we mainly focus on azimuthal plane localization since elevation cues are highly individualized and datasets are sparse (see also Sec~\ref{subjective-validation}).~The task-specific localization head then maps the outputs from temporal aggregation to a one-hot class encoding.  
Note that all the layers in the blocks use BatchNorm and LeakyReLU as the activation function.

\subsection{Deep-feature distance} \label{deep-feat-dist}

The hidden embeddings resulting from the proposed framework are then {\it aggregated} to compute a deep-feature distance (Fig~\ref{model_architecture}b). 
Although the network is trained to predict the source location, we claim that the subspace of hidden layers contains additional useful information about estimating localization similarity. 
Accumulating these hidden features to compute a deep-feature distance has been shown to correlate with human perceptual judgements in some recent studies, both in machine vision~\cite{zhang2018unreasonable} and audio~\cite{manocha2020differentiable, germain2018speech}. 
They have also been shown to be effective for representation learning without the need for prior expert knowledge~\cite{zhang2018unreasonable}.
Given a $L$-layered network, we denote the $l^{th}$ hidden layer activations as $\Fmodel_l(x) \in \mathds{R}^{T_l \times B_l \times C_l}$, 
where $T_l$, $B_l$, and $C_l$ are the time resolution, frequency bands, and number of channels respectively. 
The distance between two audio recordings is then given by, 

\begin{equation}
D(x_\mathtt{1},x_\mathtt{2})=\sum_l \frac{1}{T_l C_l B_l}  ||\Fmodel_l(x_\mathtt{1})- \Fmodel_l(x_\mathtt{2})||_1.
\label{eqn:deepfeat}
\end{equation}

\subsection{Loss Functions}
\label{ssec:loss_functions}

We train the \emph{static-source} model on temporally averaged predictions, while the \emph{moving-source} model uses finer, frame-level predictions.
Hence its reasonable to expect that the latter model captures finer estimates of DOA. 
For loss function, we use the average of \emph{label-smoothed} cross-entropy loss~\cite{szegedy2016rethinking} and \emph{haversine} distance~\cite{tang2019regression}. 

Cross-entropy is the standard loss for classification learning with neural networks.~However, label-smoothing encourages small logit gaps, which prevents model over-fitting and overconfident predictions.
If $y_k$ and $p_k$ denote the target and prediction respectively ($y_k=1$ for the correct class, $0$ otherwise), 
and $\alpha$ denotes the smoothing parameters, the loss is given by, 
\begin{equation}
H(\textbf{y},\textbf{p}) = \sum_{k=1}^{K}{- (y_k(1-\alpha) + \frac{\alpha}{K}) \log(p_k)}
\label{eqn:crossent}
\end{equation}

On the other hand, \emph{haversine} distance~\cite{tang2019regression}  emphasizes a smooth, 
continuous space for DOA predictions whereas cross-entropy focuses on a discrete space.~It captures the great-circle distance between any two points on a sphere and serves as a good proxy to compute distance between the predicted and ground-truth source location. 
Let $P=(\theta_1,\phi_1)$ and $Q=(\theta_2,\phi_2)$ be the predicted and ground truth source location from our DOA model respectively, 
where $\theta_1$ and $\theta_2$ are azimuth angles, and $\phi_1$ and $\phi_2$ are the elevation angles (all in radians), then the distance between $P$ and $Q$ is:
\begin{equation}
\label{eqn:haversine}
\resizebox{\columnwidth}{!}{
$L_{(P,Q)} = 2 \arcsin{[\sqrt{\sin^{2}(\frac{\phi_1-\phi_2}{2}) + \cos{(\phi_1)}\cos{(\phi_2)}\sin^2{(\frac{\theta_1-\theta_2}{2})}}]}$
}
\end{equation}
The predicted source locations from the DOA model are obtained by computing the softmax-weighted sum of the model predictions for each frame.


\begin{table}[t!]
\vspace{-0.12in}
\centering
\resizebox{0.75\columnwidth}{!}{
\begin{tabular}{llllll}
\hlineB{4}
    \bf Datasets          & \textbf{year}   & \textbf{\#rooms} & \textbf{\#meas.}         \\ \hline
ADREAM~\cite{winter2016database}  & 2016 & 1 & 474                  \\
AIR\_1\_4~\cite{jeub2009binaural}  & 2009 & 4 & 50                  \\
BRAS~\cite{aspock2020bras}  & 2019 & 7 & 675                  \\
Huddersfield~\cite{bacila2019360}  & 2019 & 1 & 1300                  \\
Ilmenau~\cite{mittag_christina_2016_206860}  & 2016 & 3 & 8136                  \\
IoSR~\cite{francombe2016iosr}  & 2017 & 5 & 3641                  \\
Oldenburg\_IE\_BTE~\cite{kayser2009database}  & 2009 & 5 & 296                  \\
Rostock~\cite{erbes2015database}  & 2015 & 4 & 36288                  \\
TU Berlin~\cite{wierstorf2011free}  & 2011 & 4 & 9774                  \\
Salford\_BBC~\cite{satongar2014measurement}  & 2014 & 1 & 64800                  \\
Internal Dataset  & 2020 & 1 & 6                  \\
 \hlineB{4}
\end{tabular}
}
\vspace{-0.1in}
\caption{Curated BRIR datasets for training and evaluation}
\vspace{-0.2in}
\label{brir_datasets}
\end{table}

\section{Experimental Setup}
\subsection{Datasets \& Training}
\label{datasets}

Speech recordings from the TIMIT dataset~\cite{garofolo1993darpa} are used as the source for anechoic recordings. 
The \textit{static-source} experiments are carried out using a pool of 11 Binaural Room Impulse Response (BRIR) databases, 
listed in Table~\ref{brir_datasets}. 
The resulting pool contained approximately 125k BRIR pairs from 36 different rooms.
For learning \textit{moving-source} models, we used the \textit{Hearsay} binaural audio dataset~\cite{richardneural} which contains a total of 2 hours of paired mono and echoic binaural audio from 8 different speakers. 
Participants were asked to walk around a mannequin and talk (no script was used), and their position and orientation were tracked.
In addition, we also used ambisonic audio data from the DCASE 2021 Challenge~\cite{politis2020dataset}, 
which consists of 600 1-min long sound recordings of multiple sources with annotations. 
We convert ambisonic formats to binaural using the measurements from subject 2 of the ARI HRTF dataset~\cite{ari_hrtf}.

For all cases, 3-sec audio excerpts are used for training. 
Phase and magnitude spectra of a $512$-point DFT spectrogram are extracted from these excerpts (at $16$kHz sampling rate). 
~To ensure robustness of the metric against noise, we train \OURS\, with added background noise using samples from the DNS Challenge~\cite{reddy2020interspeech}, spatialized using the BRIR datasets in Table~\ref{brir_datasets}.
For learning, we use the Adam optimizer with a learning rate of $10^{−4}$ and batch size of $32$. 
The label-smoothing parameter $\alpha$ (from Eq.~\ref{eqn:crossent}) is $0.25$. 
For all cases, $80\%$ of the data is used for training and the remainder for testing. 

\subsection{Baselines}
\label{baseline}

We compare our approach to \BAMQ\,, a binaural audio quality metric proposed by Fle{\ss}ner et al.~\cite{flessner2017assessment}.
\BAMQ\ estimates the various binaural cues (ILD, ITD, and IACC) at frame-level and combines them 
using a set of learned weights to output an overall quality metric between two recordings.
Further, observe that any learning model that is trained using binaural signals as inputs can, 
in principle, be used as a {\it surrogate} to compute a distance metric.~For instance, one can compute a deep-feature distance (similar to Sec~\ref{deep-feat-dist}) between hidden layers of a pretrained deep learning model. 
~Hence, we use two state-of-the-art binaural speech separation models - \TASNET\cite{luo2019conv} and \SAGRNN\cite{tan2020sagrnn} 
to obtain such auxiliary localization metrics.~
For both these models, we compute the average of deep-feature distances across all layers except the final decoder block as alternate baselines to \BAMQ.

\begin{table*}
\vspace{-0.15in}
\centering
\resizebox{\textwidth}{!}{
 \begin{tabular}{l l c c c c c c c c c c c c c c c c c}
 \toprule
 \multirow{2}{*}{\bf Type} & \multirow{2}{*}{\bf Name} & \multicolumn{3}{c}{\bf P1 } & \multicolumn{2}{c}{\bf P1' }& \multicolumn{1}{c}{\bf P2}& 
 \multicolumn{3}{c}{\bf P3}& 
 \multicolumn{6}{c}{\bf P4} \\
 \cmidrule(lr){3-5} \cmidrule(lr){6-7} \cmidrule(lr){8-8} \cmidrule(lr){9-11} \cmidrule(lr){12-17}
 & &\bf Speech &\bf Castanets & \bf Guitar & \bf Speech & \bf Castanets & \bf Music & \bf Speech & \bf Pink Noise & \bf Guitar & \bf Pink Noise & \bf Vocals & \bf Castanets & \bf Glocken & \bf EM & \bf AM \\
 \cmidrule(lr){1-17}
\multirow{2}{*}{\bf Pre-trained}
 & {\bf \TASNET\,} & 0.65 & 0.48  & 0.20 & 0.65 & 0.35 & 0.29 & 0.19 & 0.20 & \bf 0.10  & 0.45 & 0.01 & 0.20 & 0.12 & 0.61 & 0.69\\
 & {\bf \SAGRNN\,} & 0.72 & 0.61 & 0.21 & 0.65 & 0.40 & 0.37 & 0.20 & 0.24 & 0.07 & 0.45 & 0.14  & 0.36 & 0.19  & 0.61  & 0.72   \\
 \cdashline{1-17}
 \multirow{2}{*}{\bf DOA Models}
  & {\bf static-source} & 0.89 & 0.91  & 0.85 & 0.82 & \bf 0.94 & \bf 0.45 & 0.59 & \bf 0.33 & 0.07 & \bf 0.53 & \bf 0.62 & 0.36 & 0.45 & 0.61 & 0.79  \\
  & {\bf moving-source} & \bf 0.94 & \bf 0.94  & \bf 0.94 & \bf 0.83 & \bf 0.94 & \bf 0.45 & \bf 0.69 & 0.22 & 0.06 & \bf 0.53 & 0.61 & \bf 0.42 & \bf 0.47 & \bf 0.67 & \bf 0.83 \\
  \cdashline{1-17}
 \multirow{1}{*}{\bf Baseline}
  & {\bf \BAMQ\,} & 0.03 & 0.83  & 0.09 & 0.52 & 0.77 & -0.17 & 0.42 & 0.65 & 0.08 & -0.02 & 0.36 & 0.11 & -0.05 & 0.23 & 0.18 \\
 \bottomrule
\end{tabular}
}

\vspace{-2ex}
\caption{\textbf{Subjective evaluation}: Models include: Pre-trained models, our DOA models (including \textit{static-source} and \textit{moving-source} models), and \BAMQ, as baseline. Spearman Correlation (SC). $\uparrow$ is better.}
\label{table_mos}
\vspace{-4ex}
\end{table*}

\begin{figure}[t!]
\vspace{-0.20in}
\centering
\includegraphics[width=0.86\linewidth]{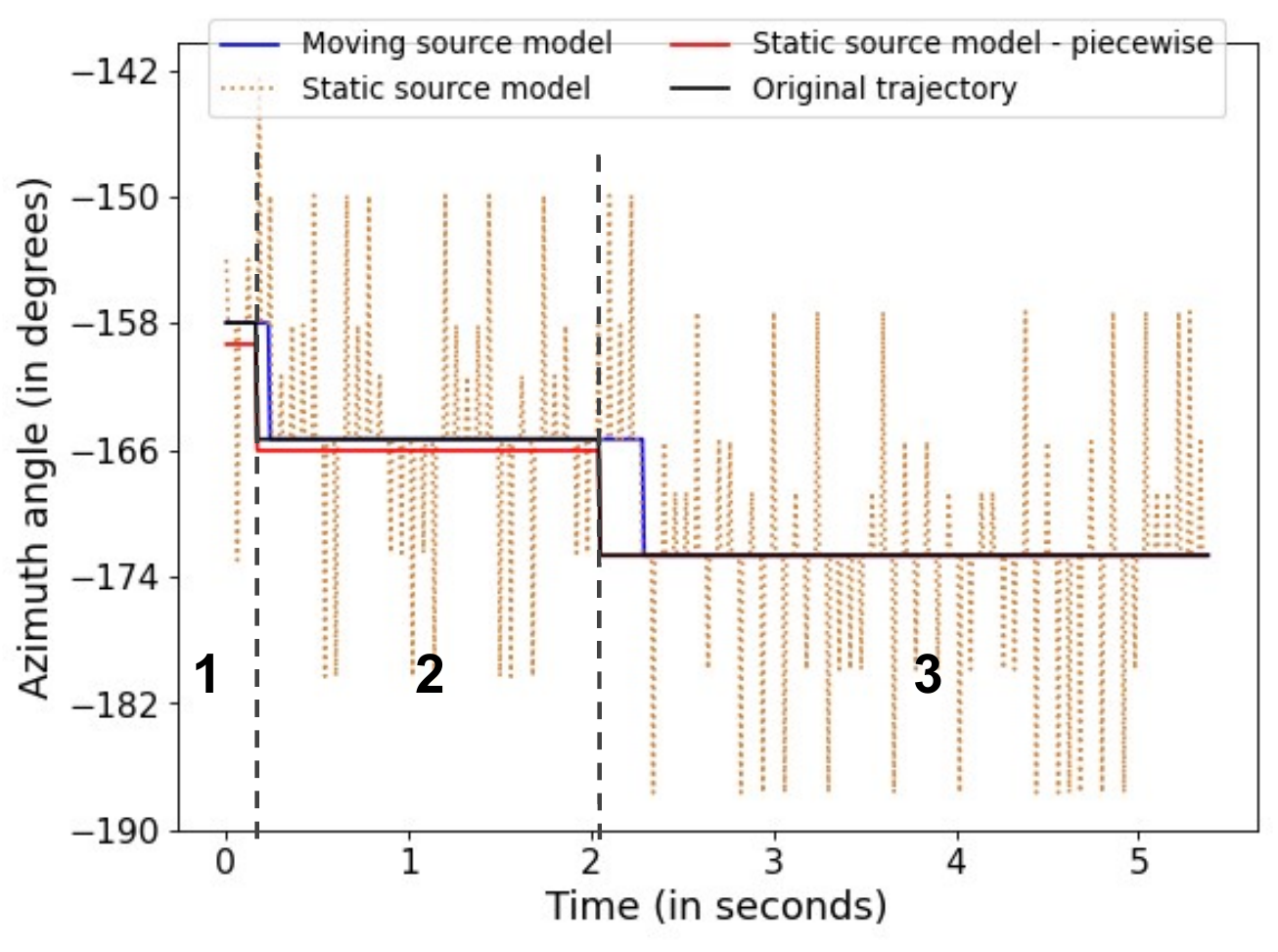}
\vspace{-0.1in}
\caption{\textbf{\emph{Framewise} localization} comparison between \emph{static} and \emph{moving} source DOA models.~The \emph{moving} trajectory in split into three intervals of \emph{constant} DOA.}
\label{compare_static_dynamic}
\end{figure}

\begin{figure}[t!]
\vspace{-0.20in}
\centering
\hspace{0.06in}
\includegraphics[width=0.84\linewidth]{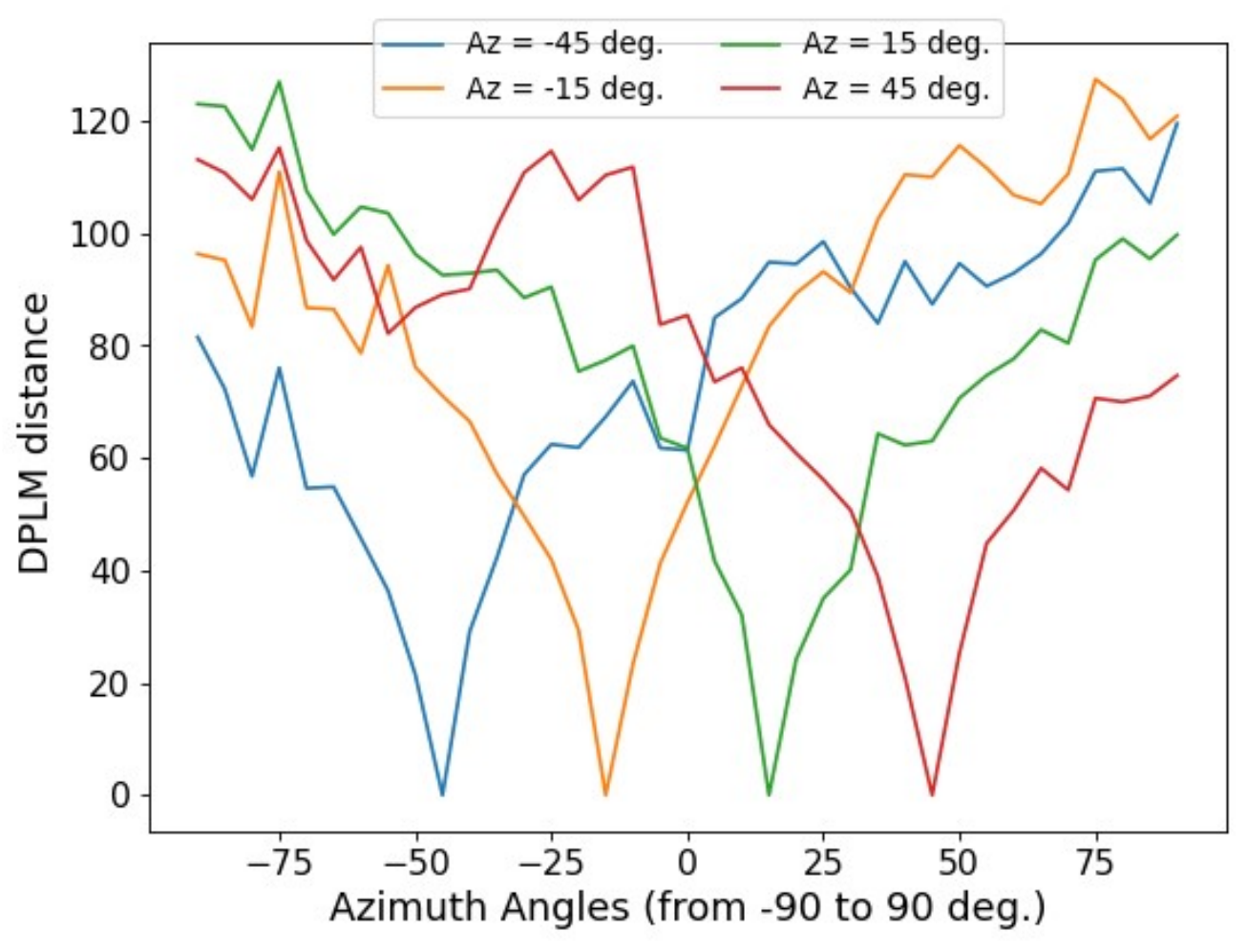}
\vspace{-0.1in}
\caption{\OURS 's variation with angular distance for four \emph{fixed} reference source positions.}
\label{same_content}
\vspace{-0.25in}
\end{figure}

\section{Results and Discussions}
\label{sec:copyright}
\subsection{Objective evaluations}

We first evaluate the \emph{static-source} and \emph{moving-source} models for localization errors on a held-out set of sound sources from TIMIT, 
spatialized using a held-out set of BRIRs.~The best performing \emph{static-source} model produced a root mean square error (RMSE) of \textbf{13.2}$^{\circ}$ in azimuth front-back folded, \emph{i.e.,} reflected about the coronal plane to discount front-back confusions).
The \emph{moving-source} model produced an RMSE of \textbf{8.4}$^{\circ}$ confirming that it leads to more accurate DOA estimates.

Fig~\ref{compare_static_dynamic} shows an example of a framewise comparison between \emph{static-source} and \emph{moving-source} models.~We observe that the framewise predictions from the \emph{moving-source} model (\emph{blue} curve) 
closely follow the actual trajectory of the source (\emph{black} curve).~
As expected, the \emph{static-source} model (\emph{orange} dotted curve) is not accurate at the frame level for tracking moving objects. 
On the other hand, the prediction improves (\emph{red} curve) when localization is computed independently for each interval, after splitting the moving trajectory into various intervals of \emph{constant} DOA (shown by the three intervals in Fig~\ref{compare_static_dynamic}).
All these observations are expected, and overall, the 
results show that the continuous temporal tracking information available to the \emph{moving-source} model helps improve the frame-level predictions, leading to fewer localization errors in general.

To verify \OURS 's sensitivity to increasing angular distance, Fig~\ref{same_content} shows the metric's distance between a \emph{fixed} reference, and a \emph{moving} test source for \emph{four} different source positions.~We see that the absolute distance values generally increase with increasing angular distance across all four source positions, indicating that \OURS\, obeys the general trend quite well. 
To quantify this trend, Table~\ref{objective_validations} shows the Spearman's rank order correlation (SC) between the output of \OURS\, and angular distance between two sources across subjects and (anechoic/echoic) listening conditions.
We see that both our models (static-source and moving-source) outperform all baselines.
Surprisingly, even the pre-trained models have a non-trivial correlation with angular distance, 
suggesting that deep-feature distances across these models serve as a good proxy for assessing localization differences between recordings.

\begin{table}[h!]
\centering
\resizebox{\columnwidth}{!}{
\begin{tabular}{llllll}
\hlineB{4}
              & \textbf{\textit{\BAMQ\,}} & \textbf{\TASNET\,} & \textbf{\SAGRNN\,}  & \textbf{\textit{static-source}} &  \textbf{\textit{moving-source}} \\ \hline
Localization  & 0.16 & 0.24 & 0.67 & 0.82 & \bf 0.86   \\
 \hlineB{4}
\end{tabular}
}
\vspace{-0.1in}
\caption{\textbf{Objective evaluation}: Correlation with \emph{angular distance}. Models include: Pre-trained models, our DOA models (including \textit{static-source} and \textit{moving-source} models), and \BAMQ\, as baseline. Spearman Correlation (SC). $\uparrow$ is better.}
\vspace{-0.25in}
\label{objective_validations}
\end{table}

\subsection{Subjective evaluations}
\label{subjective-validation}

We now use previously published diverse third-party studies 
to verify that our trained metric correlates well with subjective ratings of their tasks.~We compute the correlation between the proposed model’s predicted distance with the publicly available subjective ratings.~These correlation scores are evaluated per condition (averaging samples per condition).~The datasets used are:

\begin{enumerate}[leftmargin=0.33cm]
\vspace{-0.02in}
   \item \textbf{Bilateral Ambisonics~\cite{ben2021binaural} (P1 and P1\textquotesingle)}:
   This compares the standard and bilateral spatial audio reproduction methods across various spherical harmonic orders to assess overall quality. 
   It uses various stimuli including speech, castanets and guitar. 
   There are two versions (denoted by P1 and P1\textquotesingle), each with different subjects, 
   training sessions and different spherical-harmonic orders.
   
   \item \textbf{Spherical Microphone Array~\cite{lubeck2020perceptual} (P2)}: 
   This is designed to compare audio quality improvements across algorithms for binaural rendering of spherical microphone array 
   signals using music as stimuli. It provides an overall quality rating, with 96 variations of test signals per subject.~The pairs of recordings to be compared are not time-aligned and can be of different lengths.
   
   \item \textbf{HpEQ~\cite{engel2019effect} (P3)}: 
   This data is from headphone equalization (HpEQ) study across generic and individualized BRIRs, 
   with individualized, generic or no headphone equalization using speech, pink noise and guitar sounds for stimuli.~We have an overall quality rating, and pairs of recordings may contain very subtle differences (recordings are also not time-aligned).
   
   \item \textbf{Bitrate Compr. Ambis.~\cite{rudzki2019perceptual} (P4)}: 
   This comes from assessment of the degree of timbral distortions introduced by compression at different ambisonic orders 
   (1st, 3rd and 5th) across various modalities including simple scenes (vocals, castanets, glockenspiel, pink noise) 
   and complex scenes (EM: electronic music and AM: acoustic music).~Similar to P3, we have overall quality ratings, and recordings are not time-aligned.
\end{enumerate}

Results for the correlations with subjective ratings are in Table~\ref{table_mos}. 
Overall, our proposed metric achieves the best performance across all datasets.~Firstly, \OURS 's correlation is stable with changes in stimuli (shown by \emph{P1}, \emph{P1\textquotesingle} and \emph{P4}).~This shows the generalization power of deep-feature distance metrics, and their ability to capture attributes across speech, music, noise etc.~Furthermore, the two deep network baselines (\TASNET\, and \SAGRNN\,) trained on an unrelated task (binaural source separation) outperform \BAMQ\, on most datasets.~This suggests that deep-feature distances transfer well even across unrelated tasks, and are able to model low-level perceptual similarity well. 
However, absolute correlation values are lower for \emph{P3} showing that the metric is not robust 
enough to capture subtle differences driven by headphone equalization (some of which are very close to JNDs).
Secondly, the \textit{moving-source} model performs better than the \textit{static-source} model on most tasks, 
following a similar trend as shown in Table~\ref{objective_validations} earlier. 
Hence, frame-wise optimization for localization also seems to improve subjective ratings. 
Third, the trends with \emph{P2}, \emph{P3} and \emph{P4} suggest that 
\OURS\, (and the two deep network pre-trained baselines) are robust to non time-aligned data.
~Lastly, since \OURS\, is trained and tested under realistic, echoic, conditions, we can clearly see its 
better generalization power compared to \BAMQ\, (which is learned under anechoic environments). 

Recall that one can characterize azimuth localization by utilizing  binaural cues such as ITD and ILDs.~However, elevation localization is quite challenging because of the subject-specific influence of monaural spectral cues.
Further, lack of a wide range of elevation angles in publicly available BRIR datasets also limits building and evaluating robust models. 
We also observed similar trends in our analysis (not shown), with high error for elevation localization. 
The proposed metric performed almost the same as a simple spectral subtraction, suggesting that it does not capture elevation cues well.


\section{Conclusions and Future work}
\label{sec:pagelimit}

We present \OURS\,, a full-reference, general purpose, differentiable perceptual objective metric to assess spatial localization differences between two binaural recordings.~We show that deep-feature distances obtained from \emph{DOA} models correlate well with human ratings of localization similarity across a variety of datasets.~This is achieved without any perceptual training or calibration. In the future, we would like to extend this metric to improve elevation localization, as well as improve performance for recordings that have subtle differences.~One can also explore the utility of these differentiable metrics in deep learning based binaural speech enhancement and synthesis methods.

\bibliographystyle{IEEEtran}
\bibliography{refs21}

\end{sloppy}
\end{document}